%Paper: cond-mat/9312086
%From: zynda@guinness.ias.edu (Erika Zynda)
%Date: Tue, 21 Dec 93 13:19:08 EST
%Date (revised): Mon, 7 Feb 94 10:52:16 EST
%Date (revised): Mon, 21 Feb 94 15:54:29 EST

%Paper: cond-mat/9312086
%From: zynda@guinness.ias.edu (Erika Zynda)
%Date: Tue, 21 Dec 93 13:19:08 EST

\input phyzzx
\nonstopmode
\nopubblock
\sequentialequations
\twelvepoint
\overfullrule=0pt
\tolerance=5000
\input epsf

\line{\hfill }
\line{\hfill PUPT 1438, IASSNS 93/89}
\line{\hfill cond-mat/9312086}
\line{\hfill December 1993}
 
\titlepage
\title{Non-Fermi Liquid Fixed Point in 2+1 Dimensions}
 
\author{Chetan Nayak\foot{Research supported in part by a Fannie
and John Hertz Foundation fellowship.~~~
nayak@puhep1.princeton.edu}}
\vskip .2cm
\centerline{{\it Department of Physics }}
\centerline{{\it Joseph Henry Laboratories }}
\centerline{{\it Princeton University }}
\centerline{{\it Princeton, N.J. 08544 }}
 
\author{Frank Wilczek\foot{Research supported in part by DOE grant
DE-FG02-90ER40542.~~~WILCZEK@IASSNS.BITNET}}
\vskip.2cm
\centerline{{\it School of Natural Sciences}}
\centerline{{\it Institute for Advanced Study}}
\centerline{{\it Olden Lane}}
\centerline{{\it Princeton, N.J. 08540}}
\endpage

\abstract{We construct models of
excitations
about a Fermi surface that display calculable deviations
from Fermi liquid behavior in the low-energy limit.
They arise as a consequence of coupling to a Chern-Simons
gauge field, whose fluctations are controlled through
a ${1\over{k^x}}$ interaction. The Fermi liquid
fixed point is shown to be unstable in the infrared for $x<1$,
and an infrared-stable fixed point is found in a
$(1-x)$-expansion, analogous to the $\epsilon$-expansion of
critical phenomena.  $x=1$ corresponds to Coulomb interactions,
and in this case we find a logarithmic approach to zero coupling.
We describe the low-energy behavior of metals
in the universality class of the new fixed point,
and discuss its possible application to the
compressible $\nu={1\over2}$ quantum Hall state
and to the normal state of copper-oxide superconductors.}

\endpage

\REF\shankar {R. Shankar, Physica {\bf A177} (1991) 530; ``Renormalization
Group Approach to Interacting Fermions,'' to appear in Rev. Mod. Phys.
Jan. 1994.}
 
\REF\polchint {J. Polchinski, ``Effective Field Theory and the Fermi Surface,''
ITP prepreint NSF-ITP-92-132 (1992), to appear in the Proceedings of the
1992 TASI.}
 
\REF\ben {G. Benfatto and G. Gallavotti, J. Stat. Phys. {\bf 59} (1990) 541;
Phys. Rev. {\bf B42} (1990) 9967.}
 
\REF\haldane {F.D.M. Haldane, J. Phys. {\bf C14} (1981) 2585.}
 
\REF\wen {X.G. Wen, Phys. Rev. {\bf B42} (1990) 6623.}
 
\REF\mahan {G.D. Mahan, {\it Many-Particle Physics}, Plenum Press, New York,
1981, and references therein.}
 
\REF\affleck {I. Affleck, Nucl. Phys. {\bf B336} (1990) 517}
 
\REF\afflud {I. Affleck and A.W.W. Ludwig, Nucl. Phys. {\bf B360} (1991) 641;
Phys. Rev. Lett. {\bf 67} (1991) 161; {\bf 67} (1991) 3160;
{\bf 68} (1992) 1046.}
 
\REF\hlr {B.I. Halperin, P.A. Lee, and N. Read, Phys. Rev. {\bf B47} (1993)
7312.}
 
\REF\polchgauge {J. Polchinski, ``Low-Energy Dynamics of the Spinon-Gauge
System,'' ITP preprint NSF-ITP-93-33 (1993)}
 
\REF\varma {C.M. Varma, P.B. Littlewood, S. Schmitt-Rink, E. Abrahams, and
A.E. Ruckenstein, Phys. Rev. Lett. {\bf 63} (1989) 1996.}
 
\REF\andersongauge {G. Baskaran and P.W. Anderson, Phys. Rev. {\bf B37}
(1988) 580.}
 
\REF\andscsep {P.W. Anderson, Phys. Rev. Lett. {\bf 64} (1990) 1839}
 
\REF\gw {F. Wilczek, {\it Fractional Statistics and Anyon
Superconductivity\/} (World Scientific, Singapore 1990); M. Greiter and
F. Wilczek, Mod. Phys. Lett. {\bf B4} (1990) 1063.}
 
\REF\jain {J.K. Jain, Phys. Rev. Lett. {\bf 63} (1989) 199; Phys. Rev.
{\bf B40} (1989) 8079, {\bf B41} (1990) 7653}
 
\REF\zgw {M. Greiter, F. Wilczek, Z. Zou, paper in preparation.}
 
\chapter{Introduction}

Recent work has considerably clarified the logical foundations
of Landau's Fermi liquid theory.
Fermi liquids are now well understood, using the language of the
renormalization
group, as
interacting fermion systems whose infrared behavior
is controlled
by the free Fermi gas fixed point [\shankar,\polchint,\ben].
The quantitative success of Landau's Fermi liquid
theory is due to the fact that this renormalization
group fixed point has (almost) no relevant perturbations so long as
all interactions are local.
 
Thus generic fermion systems may be expected to
belong to this universality
class.  Several known, striking exceptions occur when
special kinematics emphasizes the infrared behavior
of
perturbations, thereby making them relevant.
One such example is the Cooper pairing instability, which is
central to the BCS theory of superconductivity.  It is
due to an interaction between
electrons of momentum ${\bf k}$ and ${\bf -k}$,
both of which
lie on the Fermi surface for any time-reversal invariant
system. This interaction is marginally relevant if attractive and
marginally irrelevant if repulsive, and causes superconductivity
in the former case. Charge- and spin-density waves are also caused
by marginal interactions present
for Fermi surfaces with special
geometries ({\it i.e}. those which exhibit nesting). These
are essentially the only known instabilities
of the Fermi liquid at weak coupling,
so that any new type of non-Fermi liquid
behavior -- {\it i.e}.
any new fixed points -- found in nature must be caused
by the presence of additional fields and/or long-range interactions.

The classification and characterization of non-Fermi liquid
metals (metals in the sense of conducting at $T=0$)
is interesting simply as a matter of principle, of course,
but gains urgency from the need to understand
the puzzling normal state
properties of the copper-oxide superconductors
and the behavior of the $\nu={1\over2}$
compressible Hall state, both of which appear to be ``almost''--
but definitely not -- Fermi liquids.

The list of established
non-Fermi liquid metals contains only two entries,
not including the BCS superconductor mentioned above --
both essentially one-dimensional -- namely the 1-D Luttinger liquid
[\haldane,\wen,\mahan] and the multi-channel Kondo effect
[\affleck,\afflud]. The first of these
is caused by the special kinematics which is the rule in
one-dimensional systems.
The interaction, which is marginal, causes the spin
and charge degrees of freedom to be separated; the retarded Green function
has a branch cut at the Fermi surface, rather than the pole
of Fermi liquid theory.
The second, the multi-channel
Kondo effect for an over-screened magnetic impurity,
is due to the presence
of an additional field (albeit one that has only a single degree
of freedom). The imaginary part of the zero-temperature
electron self-energy approaches a constant value at low frequency, and
does so as a non-trivial power of the frequency (hence, the
spectral function is non-singular at the Fermi surface).
As a result there is a non-vanishing
residual resistance at zero temperature,
which is approached as a power of the temperature;
this behavior is to be contrasted with the usual Fermi liquid
resistance which vanishes quadratically with temperature.
 
In the present work we add another
universality class to this short list of non-Fermi liquid metals,
one which is intrinsically
two-dimensional and which may be a reasonable
starting point for confronting
the theoretical challenges posed by the
aforementioned states of matter.
 
Our starting point is a system of
fermions in $2+1$ dimensions
with a Fermi surface and
low-lying excitations that interact both through
a Chern-Simons gauge field and through a $1\over{k^x}$
long-range interaction. We analyze the low-energy behavior
of this system, finding
an infrared stable fixed point
in a controlled approximation, based on the
smallness of the
parameter $(1-x)$.  This allows us to describe
some basic physical properties of these non-Fermi
liquid metals quantitatively.

Our construction of a non-Fermi liquid fixed point helps provide a
firm basis for some previous work of Halperin, Lee, and Read
[\hlr] and of Polchinski [\polchgauge].
It can also be regarded
as a concrete realization of the concept of a marginal Fermi liquid,
which was discussed from a phenomenological standpoint by Varma,
{\it et al}.
[\varma]. Let us briefly
recall some central results of these papers, which
provided much of the motivation for the present work. Halperin,
Lee, and Read considered the same model that we do, with an eye
to describing the $\nu={1\over2}$ Hall state.  In the spirit of
earlier work on the phase diagram for anyons in
a background magnetic field [\gw ] and the influential
formulations of Jain [\jain] regarding the odd-denominator
Hall states, they exploited the observation
that attaching two flux tubes of a fictitious magnetic
field to an electron -- as may be accomplished
most elegantly using
a Chern-Simons gauge field construction --
leaves it completely unchanged,
since it can be effected with a (singular) gauge transformation.
For electrons in
a spatially constant magnetic field at half-filling there
are two units of quantized
flux per electron.  Thus one can imagine attaching
two units of fictitious magnetic field to each electron in such
a way
that the total net flux -- and hence the average
field experienced by each electron -- vanishes.  In this
situation one might anticipate the possibility that
the electrons behave to a first approximation as
if there were no field at all -- and
in particular, exhibit a Fermi surface.
It is not at all obvious or trivial, however,
that it is valid to
replace the fictitious magnetic field by its average: in
fact this field
fluctuates with the density. Halperin, Lee, and Read calculated
the one-loop self-energy correction due to these gauge field
fluctuations, allowing for the $1\over {k^x}$ non-local interaction.
They found that this one-loop
self-energy exhibits non-Fermi liquid properties; specifically,
the Green function
has a branch cut at the Fermi surface, rather than a pole.
(The fermion
Green function is similar to that of the Luttinger liquid,
but there is no spin-charge separation).  However,
their self-energy calculation was uncontrolled.  The
higher-order contributions are not small; indeed, as we shall see,
it is crucial to
include radiative corrections to the fermion-gauge field vertex
in order to arrive at a consistent picture.
Polchinski considered a related system, that of fermions with a Fermi
surface interacting through a gauge field with a Maxwell action.
(We shall see that this lies in the same universality class as the
$x=0$ case of Halperin, Lee, and Read's model.)  He was
motivated by
the possibility that this model might describe the dynamics of spinons
in undoped copper-oxides. He
solved the coupled fermion and gauge field self-energy equations
self-consistently,
assuming that vertex corrections could be neglected, and thereby
obtained
results similar to those of Halperin, Lee, and Read. In order to justify
his neglect
he invoked a large $n$ approximation,
where $n$ is the number of species of spinons, that we shall argue is
not quite valid.
Varma, {\it et al}. simply assumed
a certain form for the charge- and spin-polarizability fluctuations
exchanged between electrons, and showed that they could lead
to a reasonable semi-quantitative account of anomalous behaviors in
the copper oxide superconductors.  They did not discuss the possible
origin of these fluctuations, which in our picture arise
naturally as gauge fluctuations. The one-loop
self-energy which results from the exchange of these fluctuations
leads to a Green function which, again, has a branch cut
rather than a pole at the Fermi surface.
 
In this paper, we find a non-trivial weak-coupling fixed
point for small values of $(1-x)$.
This allows us to justify lower order calculations such as those
of Halperin, Lee, and Read
and of Varma, {\it et al}.
in the regime $(1-x)\ll1$, since the
running coupling becomes small in the infrared.
We cannot shed much quantitative light on
Polchinski's results, since he considers the case $(1-x)=1$; however
one might be encouraged to think that a non-trivial
fixed point exists even in this case, as the
weak coupling fixed point might well evolve
into a strong coupling fixed point
as $(1-x)$ increases, rather than disappearing.

\bigskip
 
\chapter{Construction of the Fixed Point.}

We begin with the following effective action:
$$\eqalign{S &=
\int\,{d\omega\,{d^2}k \biggl\{\psi^{\dagger}\bigl(i\omega
- \epsilon(k)\bigr)\psi\biggr\}}
+ \int\,{d\omega\,{d^2}k\, {a_0}\epsilon_{ij}{k_i}{a_j}}\cr &\qquad
+ g \int\, {{d\omega}\,{d\omega'}\,{d^2}k\,{d^2}q
\biggl\{\psi^{\dagger}(k+q,\omega+\omega')\psi(k,\omega)
\Bigl({a_i}(q,\omega'){\partial\over{\partial{k_i}}}\epsilon(q+2k) +
a_0(q,\omega')\Bigr)\biggr\}}\cr &\qquad
+ {V_0}\int\,{d\omega}\,{d\omega'}\,{d\omega''}\,
{d^2}k\,{d^2}k'\,{d^2}k'' \psi^{\dagger}(k+k',\omega+\omega')
\psi(k',\omega')\, {1\over{k^x}}\,
\psi^{\dagger}(-k+k'',-\omega+\omega'')\psi(k'',\omega'')
\cr}\eqn\effac$$
The first two terms are the free, kinetic terms
that describe, respectively, fermionic excitations about a Fermi
surface and a Chern-Simons gauge field whose origin is left
unspecified at this time.  This Chern-Simons field is not
to be identified with the electromagnetic field.
$\epsilon(k)$,
to be further specified later, is
the single-particle energy.  It is proportional to the
Fermi velocity, $v_F$.
The next two terms are the
fermion-gauge field interaction, with coupling constant $g$,
and the non-local four-fermion interaction. In a non-relativistic
system, there is also a term of the form ${a_x}{a_x}{\psi^{\dagger}}{\psi}$
in the fermion-gauge field interaction, but this term is unimportant
according to arguments given in the appendix. We include, as well,
a term in the action of the form ${a_0}{a_0}$; such a term is allowed
by the power counting arguments below and will arise anyway as a result
of fermion loops.

The fixed point governing the low-energy
behavior of this model
will be found in a $(1-x)$-expansion,
analogous
to the $\epsilon$-expansion of critical phenomena. According
to this analogy, Fermi liquid
theory plays the role of mean field theory.
 
The ${a_0}$ equation of motion is a constraint,
$$\epsilon_{ij}{k_i}{a_j}(k) = g \int\,{d^2}q\,{d\omega'}\,
\psi^{\dagger}(k+q,\omega+\omega')\psi(q,\omega)~.\eqn\constr$$
If we substitute this constraint back into the non-local
four-fermion interaction, the important role
of this interaction becomes clear.  It takes the form
$$S_a~=~\int\,{d\omega}\,{d^2}k\,\epsilon_{ij}\epsilon_{mn}
{k_i}{k_m}k^{-x}{a_j}(k,\omega){a_n}(-k,-\omega)~.\eqn\keyterm$$
Thus as $x$ is increased, long-range fluctuations of the gauge
field are
suppressed. This is quite natural since the non-local
$1\over{k^x}$ interaction suppresses density fluctuations
(as may be seen from a calculation of the compressibility), and
the Chern-Simons term enslaves gauge field fluctuations
to density fluctuations. Note
that when $x=0$ $S_a$ is of the Maxwell form ${k_x^2}{a_y}{a_y}$
It will
turn out that for $x>1$ the gauge field fluctuations are so strongly
suppressed that the fermion-gauge field
interaction is irrelevant, and the
theory is controlled in the infrared by
the Fermi liquid fixed point. For $x<1$ the interaction
is relevant, so the stable fixed point is the
new one we construct perturbatively in $(1-x)$.
Whereas all previous non-Fermi liquids owed
their solubility to special kinematic constraints,
this model is soluble
as a result of the enslavement of transverse gauge field fluctuations
to density fluctuations, whose magnitude is directly controlled via $x$.
This possibility
is unique to two spatial dimensions, where transverse
gauge fields have only a single component.
 
As Polchinski pointed out, the significant
interaction between fermions
arises for those which are near the same point
on the Fermi surface.  To analyze this, we consider a renormalization
group
transformation that scales the system towards a single point
on the Fermi surface. Near a given point (which we will
align on the ${k_y}$-axis for simplicity) on the Fermi surface
the single-particle energy, $\epsilon(k)$, has the form:
\foot{ We emphasize
that there is nothing anisotropic here; this is simply
the form of $\epsilon(k)$ expanded to lowest order in
${k_x}$ and ${k_y}$ about
a point on the ${k_y}$-axis.}
$$\epsilon(k) = {v_F}({k_y} + a{k_x^2})\eqn\partenergy$$
(see Figure 1). Here $a=1/({2k_F})$. The correct scaling,
which scales $\epsilon(k)$ with $\omega$ and leaves
the free action invariant, is the following:
$${k_x}\rightarrow {s^{1/2}}{k_x}\eqn\xscaling$$
$${k_y}\rightarrow s{k_y}\eqn\yscaling$$
$${\omega}\rightarrow s{\omega}~.\eqn\tscaling$$
 
Under this scaling, the naive (tree-level) dimensions of the fields
and coupling constants are (${a_x}$ will be discussed in the appendix)
$$[\psi] = -{7\over4}\eqn\fscaling$$
$$[{a_y}] = -\Bigl({{7-x}\over{4}}\Bigr)\eqn\ascaling$$
$$[{v_F}] = 0\eqn\vscaling$$
$$[g] = -\Bigl({{1-x}\over{2}}\Bigr)~.\eqn\gscaling$$
Here we see, as claimed,  that $g$ is a relevant coupling
for $(1-x)>0$, so ${g^*}=0$ is no longer an infrared stable
fixed point. We will find the new fixed point, $g^*$, in an
expansion in $(1-x)$. The calculation may be done
with Wilsonian recursion relations which are closer in spirit to the
effective field theory language. In this method, recursion
relations are derived for the flow of the couplings under the
elimination of high-energy modes followed by rescalings
of the fields and coordinates.
Instead we shall proceed within the physically equivalent
field-theoretic method
more convenient for higher-order calculations, by introducing
renormalization functions $Z$, $Z_{v_F}$, $Z_g$ and calculating the
$\beta$-function.
In either case, the three diagrams
which must be calculated in order to obtain
the fixed point and scaling exponents to lowest order
in $(1-x)$ are the self-energy diagrams
and the vertex correction shown in Figure 2.
 
To be more explicit, we take the action,
$$\eqalign{S &=
\int\,{d\omega\,{d^2}k \biggl\{\psi^{\dagger}\bigl(iZ\omega
- Z{Z_{v_F}}\epsilon(k)\bigr)\psi\biggr\}}
+ \int\,{d\omega\,{d^2}k {a_0}\epsilon_{ij}{k_i}{a_j}}\cr &\qquad
+ {V_0}\int\,{d\omega}\,{d^2}k\,\epsilon_{ij}\epsilon_{mn}
{k_i}{k_m}k^{-x}{a_j}(k,\omega){a_n}(-k,-\omega)\cr &\qquad
+ {{\mu}^{{1-x}\over{2}}}
g{Z_g} \int\, {{d\omega}\,{d\omega'}\,{d^2}k\,{d^2}q
\biggl\{\psi^{\dagger}(k+q,\omega+\omega')\psi(k,\omega)
\Bigl({a_i}(q,\omega'){\partial\over{\partial{k_i}}}\epsilon(q+2k) +
a_0(q,\omega')\Bigr)\biggr\}}}~.\eqn\impeffac$$
where the renormalization functions $Z$, $Z_{v_F}$, and $Z_g$
relate the bare and physical ({\it i.e}. low-energy) quantities,
$${\psi_0}={Z^{1/2}}{\psi}\eqn\psibare$$
$${{v_F}\,_0}={Z_{v_F}}{v_F}\eqn\vbare$$
$${g_0}={{{\mu}^{{{1-x}\over{2}}}}g}{{Z_g}\over{ZZ_{v_F}}}
~.\eqn\gbare$$
We will adopt a regularization scheme that is analogous to dimensional
regularization. That is, $Z$, $Z_{v_F}$, and $Z_g$ are chosen
to cancel the pole parts in $(1-x)$ of the integrals corresponding to
the diagrams of figure 2.
 
The integrals are elementary (see the Appendix for details).
We obtain the renormalization group functions in terms of
the expansion parameter $\alpha = {{{g^2}{v_F}}\over{2\pi}}$
$$\beta(\alpha) = -(1-x)\,\alpha + 8{\alpha}^2 +
 O({\alpha^3})\eqn\bfcn$$
$$\eta_{v_F}(\alpha) = \beta(\alpha)
 {{\partial}\over{\partial \alpha}} \ln{Z_{v_F}}
= 4\alpha + O({\alpha^2})~.\eqn\anomdim$$
$ZZ_{v_F}=1$ since the one-loop self energy depends only on the
frequency and not on the momentum. As a result, $\eta = -\eta_{v_F}$.
The gauge field does not recieve any anomalous dimension at one loop
because this diagram gives a contribution $\sim {k^2}$ which is
subleading compared to $k^{2-x}$ in the infrared; the operator
dimensionality of the gauge field is, as we noted earlier, controlled
at tree level by varying $x$.
 
The fixed point ({\it i.e}. zero of the
$\beta$-function) occurs at
$${\alpha^*} = {1\over8}(1-x) + O((1-x)^2)\eqn\fp$$
and hence,
$$\eta_{v_F}({\alpha^*}) = {1\over2}(1-x) + O\bigl((1-x)^2\bigr)~.
\eqn\fpanomdim$$
Note that for $n$ species of fermions (with an explicit factor
of $n$ in front of the gauge-field kinetic term), the $\beta$-function
is:
$$\beta(\alpha) = -(1-x)\,\alpha + {8\over n}{\alpha}^2
+ O({\alpha^3})\eqn\nbfcn$$
and the fixed point is at
$${\alpha^*} = {n\over8}(1-x) + O((1-x)^2)~.\eqn\nfp$$
For large $n$ the fixed point occurs at large coupling;
all terms in the $\beta$-function are $O(n)$, so it
cannot be truncated at finite order.

At $(1-x)=0$ (Coulomb
interaction!), the fermion-gauge
field interaction is marginal and Fermi liquid theory
is approached logarithmically.  As $(1-x)$ is increased from zero,
the interaction becomes relevant -- just as,
in the case of critical phenomena, the
$\phi^4$ interaction is marginal in $d=4$ and relevant for
$\epsilon=4-d>0$. For $(1-x)$ small,
we can find a new stable fixed point at weak coupling. As we
noted above, one would expect a new fixed point, even for
$(1-x)$ not very small. None of
its properties are reliably given to low order in $(1-x)$, but
one might hope that a higher-order calculation combined with
Borel summation techniques might prove successful, as it has
in computations of critical exponents in three dimensions.
 
A basic property we wish to calculate at the new fixed point
is the time rescaling which corresponds to a given spatial
rescaling, or, simply, the anomalous dimension of the Fermi velocity.
Many of the simple physical properties of the fixed point follow from
the value of this anomalous dimension.   In particular, the Green
function aquires a branch cut when it is non-zero.

Important results may be obtained through study of the fermion
2-point function,
$$G(\omega,r) = G(\omega,{v_F}(\mu)r,\alpha(\mu),\mu) \equiv
\langle\psi^{\dagger}(k,\omega)\psi(k,\omega)\rangle~.\eqn\defG$$
Here $\mu$ is the energy scale, $r={k_y}+a{k_x^2}$ encodes the momentum
dependence of $\epsilon(k)$ near the point under consideration,
and $\alpha(\mu)$ is the coupling constant at the scale $\mu$. Rescaling
by $\mu$,
$$G(\omega,{v_F}(\mu)r,\alpha(\mu),\mu) = {\mu}^{-1-\eta}
G\bigl({{\omega}\over{\mu}},
{{{v_F}(1)r}\over{{\mu}^{1-\eta_{v_F}}}},\alpha(\mu),1\bigr)~.\eqn\scaledG$$
Taking $\mu=\omega$,
$$G(\omega,{v_F}r,\alpha(\omega),\omega) =  {\omega}^{-1-\eta}
G\bigl(1,{{{v_F}r}\over{{\omega}^{1-\eta_{v_F}}}},\alpha(\omega),1\bigr)~.
\eqn\wscaledG$$
At low energy, $\alpha(\omega)\rightarrow{\alpha^*}$, so
$$G(\omega,{v_F}r) =  {\omega}^{-1+\eta_{v_F}}
G\bigl(1,{{{v_F}r}\over{{\omega}^{1-\eta_{v_F}}}},{\alpha^*},1\bigr)~,
\eqn\Gscalform$$
where we have
substituted
$\eta_{v_F}=-\eta$ into this last equation.
The phase of the self-energy may also be obtained from
the one-loop diagram of Figure 2. When this calculation is
done with the one-loop corrected gauge field propagator
we obtain a phase
$e^{-{i\pi\over2}({{1-x}\over{3-x}})}$. Combining
these observations,
we finally arrive
at the following form for the fermion Green function:
$${\langle\psi^{\dagger}(k,\omega)\psi(k,\omega)\rangle}\, \sim \,
{1\over{{e^{-{i\pi\over2}({{1-x}\over{3-x}})}}{{(i \omega)}^{1-\eta_{v_F}}}
- \epsilon(k)}}~.\eqn\fullprop$$

This form for the Green function implies the existence of a
branch cut rather than a pole at the Fermi surface, so long as
$\eta_{v_F}>0$. Said differently, the quasi-particle weight vanishes,
$$Z \sim \lim_{{\omega}\to0} {\omega}^{\eta_{v_F}} = 0~.\eqn\qpweight$$
Here $Z$ can be defined alternatively as
$\bigl(1 - {{\partial}\over{\partial\omega}} Re\Sigma\bigr)^{-1}$,
where $\Sigma$ is the self-energy,
or simply as the wavefunction renormalization
of the $\psi$ field at the Fermi surface.
 
For the marginal case
$(1-x)=0$ -- which is, remarkably, the case of Coulomb interactions
between fermions -- there are only
logarithmic corrections to the mean-field
Fermi liquid behavior, just as arise
for critical phenomena in four dimensions or for the ultraviolet
behavior of QCD. The $\beta$-function
equation may be integrated,
$$\beta(\alpha) = 4{\alpha^2} \Longrightarrow
\alpha(\mu) \sim {1\over{8\ln\mu}}~.\eqn\mbfcnint$$
As a result the Fermi velocity recieves logarithmic scaling corrections:
$$G(\omega,{v_F}r) = {1\over{\omega{(\ln\omega)^{1/2}}}}
G\Bigl({{v_F}r\over{\omega{(\ln\omega)^{1/2}}}}\Bigr)~.\eqn\margscalform$$
Note that this one-loop renormalization group
calculation sums the leading logarithms
in {\it all\/} orders of perturbation theory.
 
\bigskip
 
\chapter{Physical Properties at the New Fixed Point}

With the scaling exponents of this theory -- and the fermion
Green function, in particular -- in hand, we may extract the basic
physical properties of these metals.  In this analysis
we follow closely the analysis of
Varma, {\it et al}., who described the phenomenology resulting from the
Green functions of the marginal Fermi liquid.
As we shall mention further below a fully realistic description of
the normal state of copper oxide superconductors may well involve
a more complicated model featuring spin-charge separation,
so this section is meant to be illustrative rather than definitive
for that application.
 
The resistivity,
$\rho$, is inversely proportional to the fermion mean free path,
$\rho\sim{({v_F}\tau)^{-1}}$, as may be seen from the
Kubo formula or from a simple Drude picture [\mahan].
Since $\tau^{-1} = {\rm Im}\Sigma \sim {{\omega}^{1-\eta_{v_F}}}$,
or, at
finite temperature, $\tau^{-1} \sim {{T}^{1-\eta_{v_F}}}$,
and ${v_F}\sim {{T}^{\eta_{v_F}}}$, the
temperature dependence of the resistivity is
$$\rho \sim  {{T}^{1-2\eta_{v_F}}}~.\eqn\res$$
This should be compared
with the usual Fermi liquid form, $\rho\sim{T^2}$.
 
This analysis may be a little too quick.  Strictly speaking,
the resistance always vanishes
in a translationally invariant system. A finite resistance
is obtained only when the underlying lattice -- umklapp processes
in particular -- is taken into account.  Also,
the {\it transport lifetime}, not simply the lifetime, must
be used in the formula for the resistivity.
The resistance is then given by
$$\rho \sim {1\over{{v_F}\tau}} {\Bigl({q\over{2{k_F}}}\Bigr)^2}\eqn
\corres$$
$q$ is the transverse momentum exchanged
in the relevant scattering processes.
Typically there is a compensation of errors, and the naive formula
gives the temperature dependence since $q$ is a reciprocal
lattice vector rather than a temperature dependent quantity
in an umklapp process.
In our model, scattering
is strong for $q^2\sim{\omega^{1-\eta_{v_F}}}$,
so one might be led to conclude, instead, that
$$\rho \sim  {{T}^{2-3\eta_{v_F}}}~.\eqn\vertres$$
However, umklapp processes should be responsible for
the scattering leading to electrical resistance,
as in the Fermi liquid case,
and we  expect \res\ to hold in practice.
In future work, we will attempt
to
address this problem more rigorously.

The tunneling conductance between one of these metals and a
conventional Fermi liquid metal is
$$g(V) = 4{\pi}{e^2}{{\mid M\mid}^2}{N_2}(0)\int {d^2}q\,{A(k,-eV)}
~.\eqn\condform$$
where $A(k,-eV)$ is the spectral function ({\it i.e}. the
imaginary part of
the Green function) of the non-Fermi liquid, $M$ is the
tunneling matrix element (assumed to be approximately constant
near the Fermi surface) and ${N_2}(0)$ is the slowly varying density
of states of the conventional metal. Since $A(k,\omega)\sim
{{\omega}^{1-\eta_{v_F}}}$ we expect, quite generally,
$$g(V) \sim {g_0} + {\mid V\mid}^{1-\eta_{v_F}}~.\eqn\tunncond$$
 
The NMR relaxation rate, $T_1^{-1}$, is given by
$$T_1^{-1} \, \sim \, \lim_{{\omega}\to0} {T\over{\omega}}
\int {d^2}q\,\hbox{Im}\chi(q,\omega)~,\eqn\nmrform$$
where $\chi(q,\omega)$ is the spin susceptibility. The lowest order
bubble diagram for $\chi(q,\omega)$ may be calculated using the
full fermion propagator \fullprop. The result is $T_1^{-1} \sim T$ just
as in Fermi liquid theory, independent of ${\eta_{v_F}}$. However,
there are also vertex corrections which must be taken into
account. These may be obtained at low energy from the $\beta$-
function. Linearizing the $\beta$-function near the fixed point
and integrating, we obtain
$$\alpha(\omega) = {\alpha^*}\biggl(1 + \Bigl({{{\alpha_0}
- {\alpha^*}}\over{\alpha^*}}\Bigr)
\Bigl({\omega\over{\mu_0}}\Bigr)^{8\alpha^*}\biggr)~.\eqn
\gatw$$
Since the vertex goes as $\mu^{{1-x}\over2}g{v_F} \sim
\mu^{{1-x}\over2}(\alpha{v_F})^{1/2}$,
$\chi \rightarrow \chi\,\omega^{({{1-x}\over2}+{\eta_{v_F}}/2)}
\,(1 + a {\omega}^{8\alpha^*})$,
or, for $\omega < T$,
$\chi \rightarrow
\chi T^{({{1-x}\over2}+{\eta_{v_F}}/2)}\,(1 + a {T}^{8\alpha^*})$.
Therefore the Fermi liquid NMR relaxation rate is replaced
by
$$T_1^{-1}\,\sim\,T^{(1+{{1-x}\over2}+{\eta_{v_F}}/2)}\,(1 + aT^{8\alpha^*}
)~.\eqn\nmr$$
At $x=1$ the coupling is marginal and flows, instead, as in \mbfcnint,
so that
$$T_1^{-1}\,\sim\,T(\ln T)^{-3/4}~.\eqn\mnmr$$
 
A number of other properties of these metals may be obtained by standard
methods.  These, too, exhibit a characteristic pattern of
deviations
from Fermi liquid behavior.

\bigskip
 
\chapter{Remarks on Applications}

In this paper, we have explicitly constructed a non-Fermi liquid
fixed point for fermions interacting with gauge fields in two spatial
dimensions. This fixed point, which may with some
justification be called a marginal Fermi
liquid, implicitly underlies the work
of Halperin, Lee, and Read and of Polchinski. These authors have
attempted to obtain the behavior of the
model by summing certain classes of
diagrams; we have, instead, used the language and methods
of the renormalization group. By showing that the physical,
renormalized coupling -- the running coupling -- approaches an
infrared stable fixed point value which is small when
$x$ is near $1$, we have justified low-order perturbation
theory in the renormalized coupling.
Given this result, the striking
semi-quantitative success
of Halperin, Lee, and Read's theory of the $\nu={1\over2}$ metallic
Hall state becomes less surprising.  For while the
bare coupling in the model they use is $2\pi$, there is a fixed
point for the low-energy behavior in the same universality class
(specifically, the same model!)
at weak coupling.  If the
theory at $g=2\pi$ is in the basin of attraction of this fixed point,
then use of low-order perturbation theory is approximately valid at
low energy.
 
It is natural to
consider the generalization of the $\nu={1\over2}$ state
of fermions to a state of anyons in a magnetic field which
cancels their Chern-Simons mean field,
say half-fermions at filling fraction
$\nu=2$, or perhaps even bosons at filling fraction $\nu=1$! These
states should be in the universality class that we have described;
again,
although the bare coupling is large, the effective coupling
at low energies is not. Experimental studies
of the $\nu={1\over2}$ state -- or one of these more exotic systems
if they can be realized in the laboratory or numerically --
should uncover the characteristic physical properties of this
non-Fermi liquid universality class, including the logarithmic
approach to free behavior.

Another possible application of this universality class is to the
description of the normal state of the copper-oxide superconductors.
Anderson and collaborators have argued that the correct theory
for the copper-oxides is a theory of fermionic
spinons, bosonic holons, and gauge fields, where the
``confining'' gauge fields are present to eliminate the redundancy in the
spinon-holon description [\andersongauge].
At half-filling (no doping), the holon
spectrum has a large gap, so the low-energy theory is a theory of
spinons and gauge fields,
which presumably falls into the $x=0$ case of the
model considered here. Polchinski considered this theory and presented
evidence for the existence of a strong-coupling fixed point, which
is the $x\rightarrow0$ extrapolation of our fixed point,
as we discussed earlier.   Unfortunately these ideas, taken at face
value, lead to Bose condensation of the holons and to
a strong coupling
theory for the spinons, with phenomenological implications that
are respectively problematic and difficult to assess.
 
It is interesting to speculate, as well,
on the relevance of our fixed point at $x=1$
to the doped copper-oxides.\foot{
One might worry that the Chern-Simons term leads to $PT$-violating
effects which are not supported by light scattering experiments.
However, $PT$-violation should be more subtle to detect since the mean-field
-- which leads to $PT$ violation at tree-level -- is cancelled.
Said differently, the ground state possesses a circular Fermi surface,
which is $PT$-invariant, so $PT$-violation can
only appear in radiative corrections, which become small in the
low-energy limit.}
The Green functions at $x=1$ are so similar to those
of the marginal Fermi liquid of Varma {\it et al}.
that provides a good
phenomenological description of these substances, that one might be
led to guess that copper-oxides are really systems of electrons,
rather than just spinons,
interacting with dynamically-generated gauge fields.
 
Things may
not be quite so simple, however. As Anderson has argued forcefully,
spin-charge separation may occur in these substances -- {\it i.e}.
spinons
and holons might be independent excitations which propagate at
different velocities in the low-energy theory -- as it
does in the one-dimensional Luttinger liquid [\andscsep].
Greiter, Wilczek, and Zou [\zgw]
have put forward a line of thought
that ameliorates some of the difficulties of the earlier spinon-holon
theories, and in which Chern-Simons fields play an important role.
Their idea is that the quantum statistics of the electron separates
at the same time as its charge and spin, so that
both the spinons and the holons are half-fermions.
They further suppose that it is valid to expand  these species
around Fermions, so that each is represented by a Fermion field
interacting with a Chern-Simons field with $|g|~=~{\pi \over 2}$.
If the mean fields of these anyons are cancelled --
perhaps most plausibly by a non-zero vacuum expectation value of the
``confining'' gauge field mentioned above --
then the spinons and holons are described at low-energies
by the fixed point of this paper.  It is plausible that
the holons, being charged,  have $x=1$.
The phenomenology of the charge carriers is
therefore governed by the weak-coupling margnal Fermi liquid fixed point,
as elaborated above.  If for some reason
the Coulomb interaction
does not survive to low energies (where it would be dynamically screened),
but is completely screened by high-energy processes,
then the holons belong to the
$x=0$ universality class.

\ack We wish to thank E. Silverstein, M. Schnitzer, and Z. Zou for
helpful discussions.

\endpage

\appendix{Calculation of the Renormalization Group Functions}

The three diagrams to be calculated are shown in Figure 2. The gauge
field self-energy diagram is not logarithmically divergent at
$x=1$, so there is no wavefunction renormalization for the gauge field.
Said differently, this diagram gives a contribution,
$${g^2}{v_F^2} \int \,{{{d\epsilon}\,{d^2}p}\over{(2\pi)^3}}
\,{i\over{i\omega +i\epsilon - \epsilon(k+p)}}
\,{i\over{i\epsilon - \epsilon(p)}}\, \sim \,
{\alpha}\Bigl( {q^2} - ({\rm const}.) {{i\mid\omega\mid}\over{q}}\Bigr)
\eqn\gaugese$$
which is subleading compared to ${q^{2-x}}$.
 
The fermion self-energy diagram is:
$${{g^2}{v_F^2}\over{(2\pi)^3}} \int\, {d\epsilon}\,{dq_x}{dq_y}\,
{1\over{q^{2-x}}}\,{1\over{i\omega - i\epsilon - \epsilon(k-q)}}~.
\eqn
\fermse$$
This is the contribution resulting from the exchange of
transverse gauge bosons. There is only one of these in 2+1 dimensions,
so in Coulomb gauge (for instance) one may solve for ${a_x}$
in terms of ${a_y}$, ${a_x}= -{{q_y}\over{q_x}}{a_y}$. In the kinematic
region of interest, ${q_y}\sim {q_x^2}/{k_F}$ -- as enforced
by the pole at this value in the ${q_y}$ integral -- so the
${a_x}-{a_x}$ propagator is suppressed by a factor of
${q_x^2}/{k_F^2}$ and the ${a_y}-{a_y}$ propagator
is all that needs to be considered. The contribution from the
${a_0}-{a_0}$ and ${a_0}-{a_i}$ propagators is subleading for a similar
reason, namely factors of $\omega$ and $q$ in the numerator.
The ${dq_y}$ integral may be done
by contour integration since ${q_y}$ appears linearly in the
denominator of the the fermion propagator. Then $\epsilon$ disappears
from the integrand, and the $d\epsilon$ integral may be done, leaving
$$2\omega\,{\alpha} \int {{dq_x}\over{q_x^{2-x}}}\,=\,
4\omega\,{\alpha}\Bigl({1\over{1-x}}\Bigr) + {\rm finite\, part}
\eqn\divfse$$
where the divergent part of the integral has been evaluated
by taking the pole part in $(1-x)$ in analogy with
dimensional regularazation. The $\omega$ integral is actually
not quite well defined, but if we use the one-loop corrected
gauge-field propagator, this is remedied, with the same
result for the divergent piece, \divfse. Since the self-energy contribution
depends only on $\omega$, we may conclude that $ZZ_{v_F} = 1$
and:
$$Z = Z_{v_F}^{-1} =
1 + 4{\alpha}\Bigl({1\over{1-x}}\Bigr) + O({\alpha^2})~.
\eqn\cterm$$
 
Finally, we turn to the vertex correction:
$${(gv_F)^2}\int{{{d\epsilon}\,{d^2}k}\over{(2\pi)^3}}\,
{1\over{i{\omega_1} + i\epsilon - \epsilon({p_1}+k)}}
\,{1\over{i{\omega_2} - i\epsilon - \epsilon({p_2}-k)}}\,{1\over{k^{2-x}}}
\,=\, {\alpha} \int {{dk_x}\over{k_x^{2-x}}}~,\eqn
\vertcorr$$
where the ${dk_x}$ and $d\epsilon$ integrals have been done
as in the self-energy integral. Again, the renormalization
counterterm is chosen to cancel the pole part in $(1-x)$,
$${Z_g} = 1 + 2 {\alpha}\Bigl({1\over{1-x}}\Bigr) + O({\alpha^2})
~.\eqn\gcterm$$
This yields a $\beta$-function,
$$\eqalign{\beta(\alpha) &=  -(1-x)
\biggl({{\partial}\over{\partial \alpha}}
\ln(\alpha{Z_g}^2/{Z_{v_F}})\biggr)^{-1}\cr
&=  -(1-x)\,\alpha + 8{\alpha}^2
+ O({\alpha^3})\cr}\eqn\abfcn$$
and hence
$$\eta_{v_F}(\alpha) = \beta(\alpha)
 {{\partial}\over{\partial \alpha}} \ln{Z_{v_F}}
= 4\alpha + O({\alpha^2})~.\eqn\aanomdim$$
 
\endpage
 
\refout
 
\endpage
 
\epsfysize=.6\vsize\hskip2cm
\vbox to .6\vsize{\epsffile{f891.eps}}\nextline\hskip-1cm
\endpage
\epsfysize=.8\vsize\hskip2cm
\vbox to .8\vsize{\epsffile{f892.eps}}\nextline\hskip-1cm

\end